\documentstyle[12pt,epsf]{article}
\begin{document}
%\draft
\sloppy
%\scrollmode
%\setlength{\unitlength}{0.02mm}
%\makeatletter
%\def\Thicklines{\let\@linefnt\tenlnw \let\@circlefnt\tencircw
%\@wholewidth4\fontdimen8\tenln \@halfwidth .5\@wholewidth}

%\newcommand{\lfrac}[2]{#1/#2}
\renewcommand{\baselinestretch}{1.3}       %
\newcommand{\deltabar}{\,\,{\bar{}\hspace{1pt}\! \!\delta }}
\newcommand{\sla}[1]{{\hspace{1pt}/\!\!\!\hspace{-.5pt}#1\,\,\,}\!\!}
\newcommand{\db}{\,\,{\bar {}\!\!d}\!\,\hspace{0.5pt}}
\newcommand{\partb}{\,\,{\bar {}\!\!\!\partial}\!\,\hspace{0.5pt}}
\newcommand{\dsla}{\partb}
\newcommand{\eql}{e _{q \leftarrow x}}
\newcommand{\eqr}{e _{q \rightarrow x}}
\newcommand{\ite}{\int^{t}_{t_1}}
\newcommand{\itz}{\int^{t_2}_{t_1}}
\newcommand{\itd}{\int^{t_2}_{t}}
\newcommand{\lfrac}[2]{{#1}/{#2}}
\title{~\\[-4cm] Anholonomic Transformations of Mechanical Action Principle}
\author{{\normalsize P.~Fiziev}
     \thanks{ Work supported by the Commission of the
       European Communities in Science and Technology,
        Contract~No. ERB3510PL920264,
       by the Sofia University Foundation for Scientific Reaserches,
        Contract~No.3052-230/94, and
       by the International Atomic Energy Agency,
       UNESCO and International Centre for
       Theoretical Physics, Trieste, Italy.}
  \\  \footnotesize International Centre for Theoretical Physics,
   \footnotesize  Trieste, Italy \\   \footnotesize  and \\
  \footnotesize Department of Theoretical Physics,
  Faculty of Physics,   \footnotesize Sofia University,\\
  \footnotesize Boulevard~5 James Boucher,    Sofia~1126,   \\
  \footnotesize Bulgaria %}
 %\author{
  \\~\\ \normalsize H.~Kleinert\\
%\address{
  \footnotesize Institut f\"{u}r Theoretische Physik,
   \footnotesize        Freie Universit\"{a}t Berlin\\
     \footnotesize      Arnimallee 14, D - 14195 Berlin
}
%%%%%%%%%%%%%%%%%%
\maketitle
%%%%%%%%%%%%%%%%%%%
\begin{abstract}
%%%%%%%
{\footnotesize
We exhibit the transformation properties of the mechanical
action principle under anholonomic transformations.
Using the fact that
spaces with torsion can be produced
by anholonomic transformations we derive the correct
 action principle in these spaces which is quite
different from the conventional action principle.
}
\end{abstract}
%~\\~\\~\\
%\pacs{03.20.+i\\ 04.20.Fy\\ 02.40.+m}
%%%%%%%%%%%%
%\maketitle
\newpage
 \section{Introduction}
 It is well known that the action formalism of classical mechanics
  is not invariant under anholonomic transformations.
When transforming the equations of motion anholonomically to new
coordinates, the result does not agree with the naively derived
equations of motion of the anholonomically transformed action
\cite{1}.
The simplest example is the free motion of a particle
with unit mass in a three-dimensional euclidean space
 ${\cal M}^3 \{ {\bf x} \} = {\cal R}^ 3 \owns {\bf x} = \{ x^i\}_{i= 1,2,3}$.
 For an orbit ${\bf x}(t)$ with velocity $\dot{{\bf x}}(t)$, the
mechanical action of the particle is
%
%
%\begin{equation}
${\cal A} [{\bf x}(t)]: = \int^{t_2}_{t_1} \frac{1}{2}
   \vert \dot{\bf x} (t) \vert^2 dt$
%\label{}\end{equation}
  %
and the Hamilton action principle:
\begin{equation}
     \delta {\cal A} [{\bf x}(t) ] = 0
\label{AP}\end{equation}
 leads to the dynamical equations
\begin{equation}
       \ddot {\bf x} = 0,
\label{DE1}\end{equation}
   solved by a straight line with uniform velocity.

Let us perform the following {\em anholonomic\/} transformation
\cite{1}--\cite{5}:
\begin{equation}
     \dot x^i(t) = e^i{}_\mu ({\bf q}(t)) \dot q^\mu (t)
\label{AT}\end{equation}
from the Cartesian  to some new coordinates
${\bf q} = \{q^ \mu \}_{ \mu =1,2,3;}$, where $e^i_ \mu ({\bf q})$
are elements of some nonsingular $ 3 \times 3$ matrix
 $ e ({\bf q}) = \parallel e^i{}_ \mu ({\bf q})\parallel$
  with $\det [e({\bf q})] \neq 0$.
  By assumption, they are defined at each point of the space ${\cal M}^3 \{
{\bf q} \} \owns {\bf q}$
  and satisfy the {\em anholonomy\/} condition
\begin{equation}
     \partial _{[ \nu }  e^i{}_ {\mu]}({\bf q}) \neq 0.
\label{AC}\end{equation}
%
%
%\subsection{}
When inserted into  the dynamical equations  (\ref{DE1})  the transformation
 (\ref{AT}) gives

%
%\begin{equation}
$$
 0 = \ddot x^i(t) = {d \over dt}\left( e^i{}_ \mu  ({\bf q}(t)) \dot q^\mu (t)
\right )
     = e^i{}_ \mu ({\bf q}(t)) \ddot q ^\mu (t)+  \partial _ \nu  e^i{}_ \mu
       ({\bf q}(t)) \dot q ^\nu (t) \dot q^\mu (t) ,
$$
%\end{equation}
%
   or, after multiplying by the inverse matrix
   $e^\alpha{}_i  ({\bf q}(t))$:
\begin{equation}
     \ddot q ^ \lambda   +  \Gamma _{ \mu  \nu }{}^  \lambda
	\dot q ^\mu \dot q^ \nu = 0.
\label{DE2}\end{equation}
These dynamical equations show that in the space
$ {\cal M}^3 \{ {\bf q ; \Gamma} \}$
 the trajectories of the particle are {\em autoparallels\/}.
Here $  \Gamma _{ \mu \nu }{}^\alpha ({\bf q}): = e^\alpha{}_i
   \partial _ \mu e^i{}_ \nu $ are the coefficients of the affine
 flat connection with zero Cartan curvature and nonzero torsion
 $S_{ \mu  \nu }{} ^  \lambda  ({\bf q}): =  \Gamma _{[ \mu  \nu ]}{}^ \lambda
   ({\bf q}) \neq 0$ (because of the anholonomic condition
  (\ref{AC})).
A different result is obtained when transforming the action
 nonholonomically. Under the transformation (\ref{AT}) it goes over into
%
%
%\begin{equation}
  ${\cal A} [{\bf q}] = \int^{t_2}_{t_1} \frac{1}{2}
     g_{ \mu  \nu } ({\bf q}) \dot{q} ^\mu \dot q^ \nu  dt,$
%\label{}\end{equation}
%
  where $g_{ \mu  \nu } ({\bf q}) := \sum_i e^i{}_ \mu ({\bf q}) e^i{}_ \nu
({\bf
 q})$
is the metric tensor in ${\cal M}^3\{ {\bf q} \}$  induced by the euclidean
metric in ${\cal M}^3\{ {\bf x}\}$ by the anholonomic transformation
(\ref{AT}). Applying variational principle in the ${\cal M}^3\{{\bf  q; g} \}$
space,
\begin{equation}
    \delta {\cal A}[{\bf q}(t) ] = 0,
\label{AP'}\end{equation}
 produces an equation of motion:
\begin{equation}
   \ddot q ^ \lambda    +  \bar  \Gamma _{ \mu  \nu } {}^ \lambda
      \dot q ^ \mu \dot q^ \nu  = 0
\label{DE3}\end{equation}
where $\bar  \Gamma _{ \mu  \nu }{} ^ \lambda  : = g^{   \lambda  \kappa }
  \bar  \Gamma _{ \mu  \nu  \kappa  },
   \bar  \Gamma _{  \mu  \nu  \lambda }: = \frac{1}{2}
   \left( \partial _ \mu g_{ \nu  \lambda   } +
    \partial _ \nu  g_{ \mu  \lambda  } - \partial _ \lambda
	 g_{ \mu  \nu }\right)$
are the  coefficients of the symmetric Levi-Cevita connection
 with zero torsion $\bar S_{ \mu  \nu }{}^ \lambda ({\bf q}) : =
    \bar  \Gamma _{[ \mu  \nu ]}{}^ \lambda  ({\bf q})
\equiv 0$.
The equations of motion (\ref{DE3}) imply that the trajectories
  of the particle are {\em geodesics\/} in the Riemannian
space ${\cal M}^3\{ {\bf  q; g} \}$.
For anholonomic coordinate transformations
these results
contradict each other
because of  (\ref{AC}).
 It is well-known from many physical examples, that
the correct equation of motion in the space ${\cal M}^3\{ {\bf q; \Gamma ,
g}\}$
are the equations (\ref{DE2}) --- the true particle trajectories
 are {\em autoparallel\/}. Hence, something is wrong with
 the variational principle (\ref{AP'}) in the space
${\cal M}^3\{ {\bf q; \Gamma,g} \}$ naturaly endowed by a Riemannian metric
${\bf g}$,
and by a nonmetric affine connection ${\bf \Gamma}$.

The purpose of this article is to show how to resolve
this conflict by an appropriate  correct
modification of  the action principle for anholonomic
coordinates. The result shows that the right
action principle must be based on the affine geometry
given by the connection  ${\bf \Gamma}$,
not on the Riemanian geometry given by the metric $\bf g$.
It has implications on the classical
mechanics in spaces with torsion calling
for a  revision of many previous publications
on the subject (See,for examle, \cite{6}, \cite{7}, and the references herein).

\section{Properties of the two tangent mappings
 $e_{q \rightarrow x}$ and $ e_{q \leftarrow x}$
}
 Consider the two $n$-dimensional manifolds ${\cal M}^n \{ {\bf  q} \}$ and
  ${\cal M}^n \{ {\bf x}\} $ with some local coordinates $ {\bf q} =
   \{ q ^\mu\}_{\mu = 1, \dots , n} $ and  ${\bf x}  = \{ x^i\}_{i = 1, \dots ,
n}$.
We shall call the space ${\cal M}^n\{ {\bf q} \}$ {\em a holonomic space\/},
  and the space ${\cal M}^n \{ {\bf x}\}$ an {\em anholonomic space\/}.
A    {\em reference system\/} on ${\cal M}^n\{ {\bf q} \}$
is defined by a local
frame consisting of $n$ linearly independent basis vector fields
${\bf e}^i ({\bf q})$ on ${\cal M}^n\{ {\bf q} \}$. They are
 specified by their components $e^i{}_ \mu ({\bf q})$.
These are supposed to form a nonsingular $n \times n$ matrix
  $e({\bf q}) = \parallel e^i{}_ \mu ({\bf q}) \parallel$ with
 $\det \left ( e ({\bf q}) \right ) \neq 0.$ Thus, there exists a set of
conjugate
 basic vector fields the components of which form the inverce matrix
$e^{-1} ({\bf q}) = \parallel e^ \mu {}_i
   ({\bf q})\parallel$. The matrix elements satisfy
   $ e^i{}_ \mu e^ \mu{}_j  =  \delta ^i{}_j$, $ e^\mu{}_i
	e^i{}_ \nu =  \delta ^ \mu {}_ \nu $. We now define
    the tangent map:
$$
      e_{q \rightarrow x}: {\hskip 1truecm} T_{\bf q}{\cal M}^n\{ {\bf q} \}
             {\rightarrow}                             T_{\bf x}{\cal M}^n \{
{\bf x} \} :
$$
by the relations :
\begin{equation}
  \left\{
\begin{array}{lll}
      \db x^i  & = & e^i{}_\mu ({\bf q}) dq^ \mu \\
       \partb_i    & = & e^ \mu {}_i ({\bf q}) \partial_ \mu \, ,
 \end{array}\right.
\label{}\end{equation}
and the inverse map
$$
     e_{q \leftarrow x}: {\hskip 1truecm} T_{\bf x}{\cal M}^n \{{\bf  x}\}
            {\rightarrow}                           T_{\bf q}{\cal M}^n\{ {\bf
q} \}
$$
by the inverse relations:
\begin{equation}
  \left\{
\begin{array}{lll}
      d q^ \mu   & = & e^ \mu{}_i  ({\bf q}) \db x^i \\
      \partial_ \mu  & = & e^i {}_ \mu  ({\bf q})\, \dsla_i \, .
 \end{array}\right.
\label{}\end{equation}

 Here $T_{\bf x}{\cal M}^n\{ {\bf x}\} $ and $T_{\bf q}{\cal M}^n \{{\bf  x}
\}$ are the
 linear tangent spaces above the points ${\bf x} \in {\cal M}^n \{{\bf  x} \} $
and
   ${\bf q} \in {\cal M}^n\{ {\bf  q} \}$, respectively. They are linearly
 transformed by the matrices ${\bf e} ({\bf q}) $ and ${\bf e}^{-1}({\bf q})$,
 which depend only on the points ${\bf q} \in {\cal M}^n\{ {\bf  q} \}$.
As a consequence of this asymmetry
there are important differences between the
basic properties of the two mappings.

In the literature \cite{1}--\cite{5}, the  mappings
 $e_{q \rightarrow x} $ and $ e_{q \leftarrow x}$
have appeared in various forms.
  The mapping $e_{q \rightarrow x}$ is called
 an anholonomic (noncoordinate) transformation if the one-forms
 $\db x^i$ are not exact: $d (\db x^i) =  \Omega _{ij}{}^k
   \db x^i  \wedge \db x^j \neq 0$, $ [ \, \dsla_i, \dsla_j] =
  -2  \Omega _{ij}{}^k \, \dsla_k \neq 0$.
The tensor $  \Omega _{ij}{}^k ({\bf q}) = e^\mu{}_i  e^\nu{}_j
\partial_{[ \mu } e^k{}_{ \nu ]} = -e^k{}_  \lambda \,  \partb_{[i}
   e^\lambda{}_{j]} $ is the so-called {\em object of  anholonomy\/}.
 For $ \Omega  _{ij}{}^k \equiv 0$,
 the coordinate transformation becomes
 holonomic. In this case one can find coordinate functions
 $x^i({\bf q})$ on ${\cal M}^n\{ {\bf  q} \}$ so that $e^i{}_ \mu ({\bf q}) =
 \partial_ \mu
x^i({\bf q})$. For $ \Omega _{ij}{}^k \neq 0$, no such functions
 exist, and the ${\cal M}^n\{ {\bf  q} \}$-space coordinates $\{q^\mu\}$
can not be treated as a true coordinates in the space ${\cal M}^n\{ {\bf  x}
\}$.
Nevertheles, they are convinient for solving dynamical equations in the last
space.

In the case of the  mapping $e_{q \leftarrow x}$, the basics
 vector fields with components $e^\alpha{}_i ({\bf q})$
define an affine geometry on ${\cal M}^n\{ {\bf  q} \}$, and it becomes an
affine space
${\cal M}^n\{ {\bf  q; \Gamma} \}$. By definition,
these vector fields are covariantly constant, i.e.,
$
  \nabla _  \mu  e ^ \lambda{}_i  =  \partial _ \mu  e ^ \lambda{}_i
     +  \Gamma _{  \mu  \nu  }{}  ^ \lambda  e^\nu{}_i  =0.
$
 This determines the coefficients of the corresponding
 affine connection to be
$
   \Gamma _{  \mu  \nu  }{}^  \lambda
     = e^ \lambda{}_k \,   \partial _ \mu
	 e^k{} _\nu .
$
This is an affine connection with {\em torsion tensor\/}
$
 S_{  \mu  \nu  }{}^  \lambda  ({\bf q}):
   =  \Gamma _{[  \mu  \nu  ]}^ \lambda ({\bf q}).
$
The torsion tensor is nonzero in the anholonomic case where
 $d(\db x^i) = S_{jk}{}^i \db x^j \wedge \db x^k,
  [\, \dsla_i, \dsla_j] = -2S_{ij}{}^k \, \dsla_k$.

The affine geometry on ${\cal M}^n\{ {\bf  q;  \Gamma} \}$ induces
   an affine geometry on the space ${\cal M}^n \{{\bf  x}\}$. The connection
coefficients are related by
$
  \Gamma _{ij}{}^k = e^ \mu{}_i  e^ \nu{}_j  e^k{}_ \lambda
	  \Gamma _{  \mu  \nu  }{} ^ \lambda   +
	  e^k{}_ \mu \, \dsla_i e^ \mu {}_j .
$
 The right-hand side vanishes showing that the  space
  ${\cal M}^n \{ {\bf x; \Gamma} \}$ is an affine euclidean space.

In the anholonomic case, the two mappings $e _{q \rightarrow x}$
 and  $e _{q \leftarrow x}$  carry {\em only \/} the tangent spaces
 of the manifolds ${\cal M}^n\{ {\bf  q;  \Gamma} \} $
and ${\cal M}^n \{ {\bf x; \Gamma}\}$ into each other.
They do  not specify a  correspondence between the points
$ {\bf q}$  and  ${\bf x}$ themselves. Nevertheless, the maps
$e_{q \rightarrow x}$ and $e_{q \leftarrow x}$ can be extended
 to maps of some special classes of paths on the manifolds
${\cal M}^n\{ {\bf  q;  \Gamma} \}$ and ${\cal M}^n \{{\bf  x; \Gamma} \}$.
Consider the set of continuous,
twice differentiable paths $ \gamma _{\bf q} (t)$: $[t_1, t_2] \rightarrow
 {\cal M}^n\{ {\bf  q;  \Gamma} \},~ \gamma _{\bf x} (t): [t_1, t_2]
\rightarrow {\cal M}^n \{{\bf
 x; \Gamma} \}$.
 Let $C_{\bf q}$ and  $C_{\bf x}$ be the corresponding closed paths
  (cycles).

The tangent maps $\eqr$ and $\eql$  imply the velocity maps:
$$
  \dot x^i (t) = e^i{}_ \mu  ({\bf q} (t)) \dot q^ \mu  (t),~~
    \dot q^ \mu (t) = e^ \mu {}_i (q(t)) \dot x ^i(t) .
$$
  If {\em in addition\/} a correspondence between {\em only two\/}
points ${\bf q}_1 \in {\cal M}^n\{ {\bf  q;  \Gamma} \}$ and ${\bf x}_1\in
{\cal M}^n \{{\bf  x; \Gamma} \}$
  is specified, say $ {\bf q}_1 \rightleftharpoons {\bf x}_1$,
 then the maps  $\eqr$ and $\eql$ can be extended to the unique maps
 of the paths $ \gamma _{{\bf q q}_1}(t)$ and $ \gamma _{{\bf x x}_1} (t)$,
 starting at the points ${\bf q}_1$ and ${\bf x}_1$, respectively.
The extensions are
\begin{equation}
  \left\{ q^ \mu (t);{{\bf q}} (t_1) = {\bf q}_1 \right\}
         {\rightarrow}
    \left\{  x^i (t) = x   ^i_1  + \int ^{t}_{t_1} e^i{}_ \mu ({\bf q}(t))
  \dot q ^\mu (t) dt;~ {\bf x} (t_1) = {\bf x}_1 \right\} ,
\label{I}\end{equation}
\begin{equation}
  \left\{ x ^i (t);{\bf x}(t_1) =  {\bf x}_t \right\}
	{\rightarrow}
\left\{ q^\mu (t) = q^ \mu _1 + \int_{t_1}^{t} e ^ \mu{}_i ({\bf q}(t))
  \dot x^i (t) dt;~ {\bf q} (t_1) = {\bf q}_1 \right\} .
\label{IE}\end{equation}
Note the asymmetry between the two maps:
 In order to find $ \gamma _{{\bf x} {\bf x}_1} (t) =
\eqr ( \gamma _{ {\bf q} {\bf q}_1} (t))$
 explicitly, one has to evaluate the integral (\ref{I}).  In contrast,
 specifying $  \gamma _{ {\bf q} {\bf q}_1} (t) =
\eql ( \gamma _{{\bf x} {\bf x}_1} (t))$
 requires solving the integral equation (\ref{IE}).

Another important property of the anholonomic maps $\eqr$
 and $\eql$ is that these do not map the cycles in one space
 into the cycles in the other space: $\eqr(C_{{\bf qq}_1}) \neq C_{{\bf xx}_1}$
 and $\eql(C_{{\bf x} {\bf x}_1}) \neq C_{ {\bf q} {\bf q}_1}$.
There exists, in general, a nonzero Burgers vector:
$$
    b^i[C_{\bf q}]: = \oint_{C_{\bf q}} e^i{}_\mu  dq^ \mu  \neq 0,~~~\mbox{if}
    ~~  \Omega_{i j}{}^k \neq 0 \, ,
$$
 and
$$
  b ^\mu  [C_{\bf x}]: = \oint_{C_{\bf x}} e^{\mu}{}_i dx^i
  \neq 0,~~~\mbox{if}~~ S_{\mu \nu}{}^{\lambda} \neq 0.
$$
\section{The Variations of the Paths in the
Space ${\cal M}^n \{ {\bf q; \Gamma} \}$}
We shall now consider variations of the paths with fixed
 ends in the holonomic space ${\cal M}^n\{ {\bf  q;  \Gamma} \}$ [See Fig.~1].
\begin{figure}[tbh]
\input kleinert.te
{}~\\[-5mm]
Figs. 1 and 2: Nonholonomic Mappings
%\caption[]{Nonholonomic Mappings}
\label{}\end{figure}
This was first done by Poincare \cite{8} for group spaces ${\cal M}^n\{ {\bf
q;
 \Gamma} \}$.
Some generalization may be found in \cite{9} and \cite{10}. Here
 we give a detailed geometrical treatment of this subject.

Let $ \gamma _{\bf q}, \bar \gamma _{\bf q} \in {\cal M}^n\{ {\bf  q;  \Gamma}
\}$ be two paths
 with common ends [See Fig.1]. According to the standard
 definitions, we consider two-parametric functions
$q^ \mu (t, \epsilon ) \in C^2$ for which: $q^ \mu (t,0) =
  q ^ \mu (t), ~~q^ \mu(t, 1 ) = \bar q^ \mu (t)$, and
 $q^ \mu (t_{1,2}, \epsilon ) = q^ \mu (t_{1,2}$). Then
 the infinitesimal increment along the path is $dq^ \mu :
 =  \partial _t q ^\mu (t, \epsilon )dt$, and the variation
 of the path is $ \delta q^ \mu : = \partial _ \epsilon
   q ^\mu  (t, \epsilon )  \delta  \epsilon $, with fixed
    ends condition: $ \delta q^ \mu \vert_{t_{1,2}} = 0$.
  We call these variations``$ \delta _q$-variations", or
more explicitly
 ``${\cal M}^n\{ {\bf  q;  \Gamma} \}$-space-variations". The above definition
leads to
 the obvious commutation relation
\begin{equation}
    \delta _q (dq ^\mu) - d ( \delta _q q^ \mu )  = 0.
\label{CR1}\end{equation}
The mapping $\eqr$ brings these paths and their variations from
  the space ${\cal M}^n\{ {\bf  q;  \Gamma} \}$ to the space ${\cal M}^n \{{\bf
 x; \Gamma}\}$.
The space ${\cal M}^n \{ {\bf x; \Gamma}\}$ contains the image paths
 before
 and after the variation
$ \gamma _{\bf x} = \eqr ( \gamma _{\bf q}),~~\bar \gamma _{\bf x} = \eqr (\bar
 \gamma _{\bf q})$.
We  now distinguish the following
 variations in ${\cal M}^n ({\bf x; \Gamma})$ [see Fig.~1]: the total
 $ \delta _q$-variation of the coordinate $x^k$: $ \delta _q x^k (t):=
  \delta _q [ x^k (t_1) + \int^{t}_{t_1} e^k{}_ \mu
   dq  ^\mu ]$, the ``holonomic variation" of the
coordinates $x^k: \deltabar x^k (t):= e^k{}_ \mu  \delta q ^ \mu  $,
 and the variations $ \delta _q (\db x^k): =  \delta _q (e^k{}_ \mu
   \delta q^ \mu )$. These have the following basic
   properties:
\\
1)~$ \delta  x^k (t_{1,2}) =0$,

i.e.,  the fixed end condition for {\em the holonomic\/}
 $ \delta _q$-variations in the space ${\cal M}^n\{\bf x; \Gamma\}$ (note that
{\em the total $ \delta _q$- variation does not possess
 this property\/}).
\\2)~
$   \delta _q x^k(t) = \deltabar x^k (t) +  \Delta ^k_q (t), $
where
\begin{equation}
 \Delta ^k_q (t): = 2\ite \partial _{[ \mu } e^k{}_{ \nu ]}
	  dq ^ \mu  dq^ \nu  = 2 \ite  \Omega _{ij}{}^k \db x^i
	   \db x^j
\label{}\end{equation}
is the {\em anholonomic deviation\/}. The function $ \Delta ^k_q (t)$
 describes the time-evolution of the effect of the anholonomy:
initially, $ \Delta ^k_q (t_1) = 0$. The final value
 $   \Delta ^k_q (t_2) = b^k$ is equal to the Burgers vector.
Then we derive the equation~:
\begin{equation}
   \delta _q (\db x^k) -d (\deltabar x^k) = 2  \Omega _{ij}{}^k
       \db x^i \db x^j.
\label{*}\end{equation}
This is  Poincare's relation.

3)~By combining the relation 2) and 3), we find
\begin{equation}
   \delta _q (\db x^k) - d ( \delta _q x^k) = 0
\label{}\end{equation}
  Under $\eqr$ mapping  the mechanical action
$  {\cal A} [ \gamma _{\bf q} ] = \itz L({\bf q, \dot q,} t) dt$
of mechanical system in the space ${\cal M}^n\{ {\bf  q;  \Gamma} \}$,
is mapped  into
${\cal M}^n\{ {\bf x; \Gamma}\}$ - action  ${\cal A} [ \gamma _{\bf x}] $
as follows:
$$
  {\cal A} [ \gamma _{\bf q}] {\rightarrow}
{\cal A}[ \gamma _{\bf x}] = \itz
   L ({\bf q}, e^{-1} {\bf \dot x},t) dt = \itz  \Lambda ({\bf q, \dot x}, t)dt
{}.
$$
There exists an associated {\em integral equation\/}
(\ref{IE}) for the orbits ${\bf q}(t) = {\bf q}[\gamma _{\bf x}]$.
 A more general form of the Lagrangian $ \Lambda $
  on the space ${\cal M}^n\{ {\bf q;  \Gamma} \}$ is:  $ \Lambda =
 \Lambda ({\bf q; x,\dot x},t)$, where ${\bf q} (t) ={\bf q}[ \gamma _{\bf x}]$
  is the corresponding functional of the path
$ \gamma _{\bf x}$. Then the Poincare relation (\ref{*})
and the definition for the anholonomic deviation $ \Delta ^k_q(t)$
give the following expression for the variational derivative
\begin{equation}
 \frac{  \delta {\cal A}  [\gamma _{\bf x}]} { \delta _q x^i(t)}
    = \dsla_i \Lambda +  \partial_i  \Lambda
      -\frac{d}{dt} \left(\partial _{\dot x^i}\Lambda \right)
	+ 2  \Omega _{ij}{}^k \dot x^j \left(\partial _{\dot x^k}
	     \Lambda  + \itd \partial _k  \Lambda d\tau \right)
\label{}\end{equation}
where $\dsla_i = e^ \mu{}_i   \partial _ \mu $.
Note the additional force-term proportional to the object of the anholonomy
$ \Omega_{i j}{}^k $.

The variational principle $ \delta _q {\cal A} [ \gamma _{\bf x}] = 0$
implies therefore the following generalized Poincare equations
(\ref{GPE}):
\begin{equation}
   \frac{ \delta {\cal A} [ \gamma _{\bf x}]}{  \delta _q x^i (t)} = 0;~~~
   q^ \mu (t) = q ^\mu _1 + \int^{t}_{t_1} e^ \mu {}_i (q (t))
     \dot x^i(t) dt
\label{GPE}\end{equation}
in the anholonomic space ${\cal M}^n \{{\bf x; \Gamma} \}$.
\noindent
We have to stress that:
\\
1.~In general, (\ref{GPE}) is a system of {\em integro-differential
equations\/}, instead of systems of ordinary differential equations.
\\
2.~The nonlocal term which is proportional to $\int^{t_2}_{t} \partial _k
    \Lambda d\tau $ violates causality. It may be shown that if,
   and only if, the Lagrangian $ \Lambda $ originates from some
   ${\cal M}^n\{ {\bf  q;  \Gamma} \}$-space Lagrangian $L({\bf q,\dot q}, t)$,
then
   $\partial _k  \Lambda \equiv 0$  and no causality problem appears.
\\
3.~Poincare himself considered only the special case when:
\\
i)~$\dsla_i$ are independent left invariant vector fields on a Lie
     group space. In this case, $-2 \Omega _{ij}{}^k = C_{ij}{}^k = const $
are just structure constants of the Lie-group.
\\
ii)~$L=L_0$, where $L_0$ is invariant under corresponding group
    transformations, and hence, $\dsla_k  \Lambda _0 =0$.

Under these two conditions, the (\ref{GPE}) reduces to a closed
    system of ordinary differential equations:
$$
   \frac{d}{dt} \left(\partial _{\dot x^i}  \Lambda _0\right)
     + C_{ij}{}^k \dot x^j \partial _{\dot x^k} \Lambda _0 =0.
$$
 If instead of ii) we have $L=L_0 - U({\bf q})$, then there exists
 causal system of integro-differential equations:
\begin{equation}
    \left\{ \begin{array}{l}
      \frac{d}{dt} \left(\partial _{\dot x^i} \Lambda _0\right)
	+ C_{ij}{}^k \dot x^j \partial _{\dot x^k}  \Lambda _0
	    = - \partial _i U({\bf q}),\\
               \\
	q  ^ \mu (t) = q^ \mu _1 + \int^{t}_{t_1} e ^\mu{} _i (q(t)
		) \dot x^i dt.
\end{array}               \right.
\label{PE}\end{equation}
which are called Poincare equations  \cite{10}--\cite{12}.

The most popular example where the Eqs.~(\ref{PE})
apply are the Euler equations for rigid body rotations in the
body-fixed reference system \cite{10}--\cite{12}, mentioned already
by Poincare himself \cite{8}. Let us derive these equations
and extend them to a description  of the full
rigid body dynamics (including translations)
in the body system  \cite{13}.

In this case ${\cal M}^{(6)} \{{\bf q}\} = {\cal SO}(3)  \times {\cal
R}^3,~{\bf q} =
\{ \varphi^ \mu, x ^\mu \}$, where $  \varphi ^\mu $ are
the Euler angles, and $x ^\mu$ are center-of-mass-coordinates in the
stationary system; $ \mu , i = 1,2,3$; the group constants
are $C_{ijk} = - 2  \Omega _{ijk}  =  \epsilon _{ijk}$, where
 $ \epsilon _{ijk}$, is the Levi-Cevita antisymmetric symbol.
The Lagrangian in the space ${\cal M}^{(6)} \{{\bf q}\}$ reads
$
  L = \frac{1}{2} I_{ \mu  \nu }(\varphi) \dot{\varphi}^ \mu
 \dot \varphi^ \nu + \frac{1}{2} M \delta _{  \mu  \nu  }
  \dot x^  \mu \dot x ^ \nu
$
where $I_{  \mu  \nu  }(\varphi)  =
   \sum_{i} I_i e^i{}_{ \mu } (\varphi) e^i{}_{\nu } (\varphi)$
%\vfill
are the components of the bodies inertial tensor, and
 $I_i= const$ are the principal inertia momenta. The anholonomic
 space is
${\cal M}^{(6)} \{ {\bf x} \} = {\cal M}^{6} \{ \Phi^i , X^i\}   $,
where
$\db \Phi^i = e^i{}_ \alpha ( \varphi) d\varphi^ \alpha  =  \Omega ^i dt$
(~$\Omega^i$  are the components of the angular velocity
 in the body system), and
$ \db X^i =  \epsilon ^i{} _ \alpha  (\varphi) dx^ \alpha  = V^i dt $
(~$V^i$ are the components of the center-of-mass
 velocity in the same system). Then, in the space
 ${\cal M}^{(6)}\{ \Phi^i, X^i\}$, the Lagrangian reads
$$
 \Lambda  (\dot \Phi, \dot X) = \frac{1}{2} \sum _{i}
      I_i  \dot \Phi^i \dot \Phi ^i  + \frac{1}{2} M
     \sum_{i}  \dot X^i \dot X^i .
$$
  Using the vectors of  angular momentum ${\bf K} = I {\bf   \Omega }$,
  and  momentum  ${\bf P} = M{\bf V}$,
 we can write down the variational derivatives in  the  form:
$$
   \frac{ \delta {\cal A} }{  \delta _q \Phi}  =
      \frac{d{\bf K}}{dt}  + {\bf  \Omega } \times {\bf K},~~~~
      \frac{ \delta {\cal A}}{  \delta _q {\bf X}} =
       \frac{d{\bf P}}{dt} + {\bf  \Omega } \times {\bf P}.
$$
 Hence, the variational principle $ \delta _q {\cal A}[ \gamma _{\bf x}] =0$
  gives the correct equations of motion. They contain
  gyroscopic terms
  proportional to $ \Omega :  \Omega  \times {\bf K}$
and $  \Omega \times {\bf P}$. The usual variational principle
$ \delta _x {\cal A}[ \gamma _{\bf x}] = 0$ in the anholonomic
 space ${\cal M}^{(6)} \{ \Phi^i, X^i\}$ would have produced wrong equations
 of motion without gyroscopic terms.
\\
\section{The Variations of the Paths in the Space ${\cal M}^n \{ {\bf x;
\Gamma}\}$}
Consider now the variation of the paths with fixed ends in
the anholonomic space ${\cal M}^n\{\bf x; \Gamma\}$ [See Fig.~2]. This
consideration
is similar to the one above, but it has some specific features.
Let $ \gamma _{\bf x}, \bar \gamma _{\bf x} \in {\cal M}^n\{\bf x\}$ are two
paths
with common ends [see Fig.~2]. We consider two-parametric
 functions $x^i (t,  \varepsilon ) \in C^2 $ for which:
$ x^i (t,0) = x ^i(t), x^i (t,1) = \bar x^i (t)$,
and $x^i (t_{1,2},\varepsilon) = x^i(t_{1,2})$. Then the infinitesimal
 increment along the path is $dx^i: = \partial_t x^i (t, \varepsilon)dt$,
and  the variation of the path is $ \delta x^i: = \partial_\varepsilon x^i
(t, \varepsilon )  \delta \varepsilon$, with fixed ends:
$  \delta x^i \vert_{t_{1,2}  } =0$. We call these variations
`` $\delta _x$-variations", or  more explicitly
``${\cal M}^n\{{\bf  x; \Gamma}\}$-space-variations".
 The above definitions lead to the obvious commutation relation
\begin{equation}
    \delta _x (dx^i) - d( \delta _x x  ^i ) = 0.
\label{CR3}\end{equation}
 The  $\eql$ mapping maps these paths and their variations from
  the space ${\cal M}^n \{\bf x; \Gamma\}$ to the space ${\cal M}^n\{ {\bf  q;
\Gamma} \}$.
According to our definitions, they go over into
 ${\cal M}^n\{ {\bf  q;  \Gamma} \}$-paths $ \gamma _{\bf q} = \eql ( \gamma
_{\bf x}), {\bar \gamma}_q
   = \eql (\bar  \gamma _{\bf x})$ [see Fig.~2], and we have got : the ``total
  $ \delta _x$-variation" of the coordinates $q^ \mu : \,
%%%
 \delta _x q^ \mu (t):=  \delta _x \left[  q^ \mu  (t_1) +
      \ite e^ \mu {}_i ({\bf q}) dx^i\right]$
(this is an indirect definition accomplished by integral
 equation  (\ref{IE}), the ``holonomic variation" of the coordinates
 $q ^ \mu : \deltabar_x q^ \mu (t):= e^ \mu{} _i ({\bf q})  \delta x^ i$,
and the variations $ \delta _x (dq^ \mu ): =  \delta _x
\left[ e^ \mu {}_i({\bf q}) dx^i\right] $. These have
 the following basic properties:
%%%
\\
1)~$\deltabar_x q^ \mu  (t_{1,2}) = 0$,\\
   i.e.\ the fixed end condition
   for {\em the holonomic\/} $ \delta _x$-variations in the space
   ${\cal M}^n\{ {\bf  q;  \Gamma} \}$ (note that {\em the total $ \delta
 _x$-variations
  do not possess this property\/}).
\\
2)~$ \delta _x  q ^\mu  (t) = \deltabar_x q^ \mu  (t) +  \Delta  ^\mu
       _x (t)$, \,
where \\
$$
       \Delta ^ \mu _x (t): =  \ite d\tau   \Gamma _{ \alpha  \beta }{}
     ^ \mu  \left( dq^ \alpha  \deltabar_x q ^\beta - \delta _x q ^\alpha
      dq^ \beta \right)
$$
is the {\em anholonomic deviation\/}. The function $  \Delta ^ \mu _x
  (t)$ describes the time-evolution
     of the effect of the anholonomy: $ \Delta  ^\mu_x (t_1) = 0$.
   The final value $ \Delta  ^\mu_x (t_2) = b^ \mu $ is the Burgers
vector.   \\
3)~
$
     \delta _x (dq^ \mu ) - d ( \deltabar_x q^ \mu ) =
  \Gamma _{ \alpha  \beta   }{}^ \mu  \left( dq^ \alpha  \deltabar_x q^  \beta
    -  \delta _x q ^\alpha  dq^ \beta \right).
$

 This  relation is new, replacing the Poincare relation
     (\ref{*}). It may be rewritten in a more elegant form:
\begin{equation}
    \delta _{x A } (dq^ \mu ) - d _A
       (\deltabar_x q^ \mu ) = 0.
\label{**}\end{equation}
 Here $ \delta _{x A} (d q^ \mu ) : =  \delta _x (dq^ \mu )
     +  \Gamma _{ \alpha  \beta }{}^ \mu \delta_x {q^ \alpha}  dq^  \beta $
    is the ``absolute variation" and $d_ A (\deltabar_x q^ \mu ):= d
     (\deltabar_x q^ \mu ) +  \Gamma _{ \alpha  \beta }{}^ \mu  dq^ \alpha
 \deltabar_x q^  \beta $ is the corresponding ``absolute differential".
By combining the last two equations we find
\begin{equation}
  \delta _x (dq^ \mu ) - d ( \delta _x q ^ \mu ) = 0.
\label{}\end{equation}
 For anholonomic deviation $ \Delta ^ \mu_x (t)$, we introduce the
vector notations $ \Delta _x (t) = \{  \Delta ^ \mu _x (t) \}_{ \mu =1,\dots
 n}$
 and derive an ordinary differential equation. Written
 in a matrix form and imposing the proper initial condition, it reads:
\begin{equation}
 \left\{
\begin{array}{lll}
   \dot  \Delta _x (t) &= & -G(t)  \Delta _x (t)  + S (t) \deltabar_x q(t) \,
,\\
   \\
    \Delta _x (t_1)&  =& 0 \, .
\end{array} \right.
\label{}\end{equation}
 Here $G (t) = \{ G^  \nu   {}_\mu  (t)\} :=
    \{ \Gamma _{  \mu  \lambda  } {}^  \nu   ({\bf q} (t) )
    \dot q^ \lambda  (t) \}$ and $S(t) = \{ S^\nu  {}_\mu  (t) \}:=
  \{ S_{ \lambda  \mu }{}^ \nu  ({\bf q} (t)) \dot q^ \lambda (t)\} $
  are time dependent $n \times n$-matrices. The solution of the initial
 value problem is:
\begin{equation}
    \Delta _x (t) = \ite d\tau  U(t,\tau )  S(\tau ) \deltabar_x q(\tau ),
\label{}\end{equation}
where
\begin{equation}
 U(t,\tau ) = T -\exp \left( \int^{t}_{\tau } G(\tau ')
     d\tau '\right) .
\label{}\end{equation}
 For a mechanical system in the space ${\cal M}^n \{\bf x; \Gamma \}$ with
action
$ {\cal A}[  \gamma _{\bf x}] = \itz  \Lambda ({\bf x, \dot x}, t) dt$,
the action is mapped under $\eql$ mapping as follows:
\begin{equation}
  {\cal A}[ \gamma _{\bf x}] \rightarrow {\cal A} [ \gamma _{\bf q}] =
    \itz  \Lambda ({\bf x,\dot q,} t) dt = \itz L({\bf x, q, \dot q,} t)dt \, .
\label{}\end{equation}
 The correspondence between the orbits is given by the  associate {\em integral
\/}
 (\ref{I}) for ${\bf x}(t) = {\bf x} [ \gamma_{\bf q} ]$.

A more general form of the Lagrangian $L$ on the space
 ${\cal M}^n\{ {\bf  q;  \Gamma} \}$ is $L =  L ({\bf x; q, \dot q,} t)$, where
 ${\bf x}(t) = {\bf x} [ \gamma _{\bf q}]$ is the corresponding functional
 of the path $ \gamma _{\bf q}$. Then the relation (\ref {**})
  and the definition for the anholonomic deviation $ \Delta _x(t)$
 give the following expression for the variational derivative
\begin{eqnarray}
   \frac{ \delta {\cal A} [ \gamma _{\bf q}]}{ \delta _x q^ \mu (t)  }
    & =& \dsla_ \mu L + \partial _ \mu L - \frac{d}{dt}
     \left(\partial _{ {\dot q}^ \mu} L\right) +\nonumber \\
  &&  + 2 S_{ \nu  \mu }{}^ \lambda  \dot q^ \nu  \left(
    \partial _{\dot q^ \lambda }L + \itd \left( \partial _ \sigma
      L -  \Gamma _{ \sigma   \alpha  }{}^ \beta   \dot q^\alpha
      \partial _{\dot q ^ \beta  }  L \right) U^ \sigma {}_ \lambda
	d\tau \right)
\label{}\end{eqnarray}
where $\dsla_ \mu = e^i{}_ \mu  \partial _i$.
 Observe the additional force-term proportional to the torsion $S_{ \nu  \mu }
{}^ \lambda $.

Hence, the variational principle $ \delta _x {\cal A} [ \gamma _{\bf q}] =0$
implies the following integro-differential dynamical equations :
\begin{equation}
  \frac{ \delta {\cal A } [ \gamma _{\bf q}]}{   \delta _x q^ \mu (t)}
   = 0;~~ x^i (t) = x^i_1 + \ite e^i{}_ \mu  ({\bf q}(t)) \dot q^ \mu  (t) dt
\label{DE4}\end{equation}
 in the holonomic space ${\cal M}^n\{ {\bf  q;  \Gamma} \}$.

The following  points have to be emphasized.
\\
1.~The nonlocal term which is proportional to the term
\begin{equation}
     \itd \left( \partial_  \sigma
     L -  \Gamma _{  \sigma  \alpha  }
     {}^ \beta  \dot q ^ \alpha  \partial _{\dot q^\beta }L\right)
     U^  \sigma  {}_ \lambda  d\tau
\label{}\end{equation}
violates the causality.
\\
2.~It may be shown that if, and only if, the Lagrangian $L$ originates from
 some ${\cal M}^n \{{\bf  x; \Gamma}\}$-space Lagrangian $ \Lambda ({\bf x},
\dot {\bf x}, t)$
  then $\partial _  \sigma  L-  \Gamma _{  \sigma  \alpha  }{}^ \beta
  \dot q^ \alpha  \partial _{\dot q^ \beta } L \equiv 0,$ so that no
causality problem appears. In this case,
we have the following system of nonlocal, but causal integrodifferential
 equation as  dynamical equations:
\begin{equation}
    \left\{
\begin{array}{ll}
 \dsla_ \mu L + \partial _ \mu  L - \frac{d}{dt} (\partial  _{\dot q^ \mu } L
      ) + 2 S_{ \nu  \mu }{}^ \lambda  \dot q^ \nu  \partial _{\dot q^ \lambda
}
     L = 0\nonumber \\
     \\
   x^i (t) = x^i_1 + \ite e^i{}_ \mu  ({\bf q}(t)) \dot q^ \mu  (t) dt
\end{array}\right.
\label{}\end{equation}
3.~If the ${\bf x}$-space Lagrangian has the form $ \Lambda ({\bf x},
 \dot{\bf  x}, t)
 =  \Lambda  (\dot {\bf x}, t)$,
then there exists a  complete local system of ordinary differential
equation as a dynamical equations :
\begin{equation}
     \partial _ \mu L - \frac{d}{dt} (\partial _{\dot q^ \mu }L )
       + 2 S_{ \nu  \mu }{}^ \lambda  \dot q ^ \nu
      \partial _{\dot q ^ \lambda } L = 0.
\label{DE6}\end{equation}
In these equations, an additional ``torsion force"
\begin{equation}
      {\cal F}_ \mu  = 2 S_{ \nu  \mu }{}^ \lambda  \dot q
	 ^ \nu  \partial _{\dot q^ \lambda} L = 0
\label{}\end{equation}
  appears. It is easy to show, that this force preserves the mechanical
 energy but it is a nonpotential force even in the generalized sense:
 the force cannot be written as a variational derivative of some
 suitable functional:
 $ {\cal F}_ \mu  \neq  \delta  ( \int V({\bf q, \dot q,
  \ddot q} , \dots; t) dt ] /  \delta q^\mu (t) $.

The dynamical equation in presence of torsion, generated by anholonomic
transformation, may be rewritten in a form:
\begin{equation}
  \frac{ \delta {\cal A} [ \gamma _{\bf q}]}{  \delta _x q^ \mu (t)}
     = \frac{ \delta {\cal A}[ \gamma _{\bf q}] }{ \deltabar q^ \mu (t)}
     + {\cal F}_ \mu  = 0.
\label{}\end{equation}
   The correct variational principle $ \delta _x {\cal A} [ \gamma _{\bf q}] =
0$
can be replaced by a modified D'Alembert's principle:
\begin{equation}
     \delta _q {\cal A} [ \gamma_{\bf q}] + \itz {\cal F}_ \mu   \delta q^ \mu
= 0.
\label{}\end{equation}
In this form, it involves only ${\cal M}^n\{ {\bf  q;  \Gamma} \}$-space
variables.
  \\
{\em Examples:\/}\\
1.~Consider a free particle in an affine flat space ${\cal M}^n\{ {\bf  q;
\Gamma} \}$
 with
 nonzero torsion \cite{12}. We can think this space as  produced by some
tangent mapping
$\eql$ from the euclidean space ${\cal M}^n\{ {\bf  x;  \Gamma} \}$. Then
under this mapping  $  \Lambda ({\bf \dot x}) = \frac{1}{2} m \sum_{i} (\dot
x^i)^2 \,
    {\rightarrow} \, L = \frac{1}{2} mg_{ \mu  \nu }
      ({\bf q}) \dot q ^\mu \dot q^ \nu ,~~
     g_{ \mu  \nu }  ({\bf q}) = \sum_{ i}  e^i{}_ \mu ({\bf q})  e^i{}_ \nu
({\bf q}) $,
and the variational principle $ \delta _x {\cal A} [  \gamma _{\bf q}] =0$
leads to the correct equation of motion (\ref{DE2}):
$$
  g_{  \mu  \nu }\left(\ddot q ^\nu +  \Gamma _{  \lambda  \sigma  }{} ^\nu
      \dot q^\lambda   \dot q^ \sigma  \right) =
       g_{ \mu  \nu } \left(\ddot q^ \nu  + \bar  \Gamma _{ \lambda  \sigma }
    {}^ \nu \dot q^\lambda   \dot q^\sigma  \right) +
 {\cal F}_ \mu  = 0.
$$
The torsion force reaches the conflict described in the
 introduction.
\\
2.~The Kustaanheimo-Stiefel transformation in celestial mechanics:\\
Consider the Kepler problem in the celestial mechanics. Here
${\bf x}\in{\cal R}^3~,   r=\ \vert {\bf x}\vert$, and the Lagrangian reads:
 $  \Lambda ({\bf x, \dot x}) = \frac{1}{2} m \vert {\bf \dot x} \vert^2 +
  \frac{ \alpha }{r}$ ($ \alpha  =$ const). The well-known dynamical
 equations in  ${\cal R}^3 \{{\bf  x}\}$ space are $m \ddot{\bf x} +  \alpha
{\bf x}
 /r^3 = 0$. Let us perform the Kustaanheimo-Stiefel
 transformation \cite{1}, \cite{13}, which transforms the Kepler
 problem to the harmonic oscillator problem in ${\cal R}^4 $, setting:
$$
\begin{array}{ll}
   x^1 = 2 \left(u^1 u^3 + u^2 u^4\right),
 &
\\
   x^2 = 2 \left(u^1 u^4 + u^2 u^3\right),&\\
   x^3 = \left(u^1\right)^2 + \left(u^2 \right)^2 -
	 \left(u^3\right)^2 - \left(u^4\right)^2,&
\end{array}
   ~~~\mbox{for}~~~\vec{\bf u} = \left(
\begin{array}{l}
   u^1\\
    u^2\\
     u^3\\
      u^4
\end{array}  \right)
  \in {\cal R}^4.
$$
 Note that a variable $x^4$ does not exist;
 this is {\em not\/} a coordinate
transformation. An {\em  anholonomic\/} transformation
may be defined by
 $\db x^i = e^i{}_ \mu  (\vec u) du^\mu
   ,~~i = 1,2,3,4;$ using the matrix:
$$
 e(\vec u) = \|  e^i{}_ \mu (\vec u)  \| =   \left\|
\begin{array}{cccc}
   u ^3 & u^4  & u^1 & u^2 \\
   u ^4 &  -u^3 & -u^2 & u^1 \\
   u^1  & u^2 & -u^3  & -u^4 \\
   u^2 & -u^1  & u^4  & -u^3
\end{array}            \right\|  .
$$
It is easy to check that $d ( \db  x^i )= 0$ for $i = 1,2,3;$ but
 $d  (\db x^4 ) \neq 0.$ Hence we have indeed anholonomic transformation
 of the type $ e_{u \leftarrow x}$.
The fourth (anholonomic) coordinate $x^ 4$
must be inserted into
  the original Lagrangian
$ \Lambda ({\bf x}, \dot{\bf  x})$, replacing it by a new one:
   $ \bar  \Lambda ( {\bf \vec x, \dot {\vec x} } ) =  \Lambda ({\bf x},
\dot{\bf  x})
    + \frac{1}{2} m ( \dot x ^4) ^2$. The new
   four-dimensional problem in ${\cal R}^4 \{ {\bf \vec x} \}$ will be
equivalent to the old one if one enforces the constraint $x^4 =$ const.
 The $e_{ u\leftarrow x}$ mapping maps the Lagrangian
 $ \bar  \Lambda ( {\bf \vec x, \dot {\vec x} } )$ into the
 space ${\cal R}^4 \{ {\bf \vec u }\}$, where it reads:
$ L( {\bf \vec u, \dot {\vec u} } ) = 2 m{\bf \vec u } ^2  {\bf \dot{\vec u} }
^2+
     \lfrac{ \alpha }{ {\bf \vec u}^2}$. One can produce the correct dynamical
   equations in the space ${\cal R}^4\{ {\bf \vec u} \}  $ using our
modification
  of the variational principle: Here $ \delta _x {\cal A} [{\bf \vec u} (t)]=0$
 implies the equation of motion  $ \lfrac{\delta {\cal A}}
  { \delta u^ \mu   } + {\cal F}_ \mu  = 0$ containing a torsion force
 ${\cal F}_ \mu =  m \dot x^4 \dot S_{1 \mu 2}$ (in this problem
  the only nonzero components of the torsion are $\{S_{1 \mu 2}\} =
   \{ S_{ 3 \mu 4}\} = 4 \{ u^2, -u^1, u^4, -u^3\} \, \cite{1} \, )$.
The application of the naive action principle in the space
 ${\cal R}^4 \{ {\bf \vec u} \}$ would produce a wrong equation of motion,
lacking the
 torsion force.
\\
\section{  Concluding remarks}
It is obvious, that the variational principles described
in this article  have many other applications, for example
in the $n$-body problem of celestial mechanics, in  rotating systems
 (analogously to the Euler equations for a rigid body),
in the field theory of gravitation with torsion $ \cite{6}, \cite{7} $,
 and in many other physical systems $ \cite{13} - \cite{18}$ , where
 anholonomic coordinates are a convenient and a useful tool .

One of us (PPF) would like to thank the Commission of the
European Communities   in Science and Technology for support, which permitted
him to collaborate with Professor~Hagen~Kleinert,
the Freie Universit$\ddot a$t-- Berlin, for hospitalization during the two
visits in 1993--94,
as far as  the Sofia University Foundation for Scientific Reaserches,
Professor~Abdus~Salam, the International Atomic Energy Agency, UNESCO,
and the International Centre for Theoretical Physics for support
during the visit in ICTP, Trieste in October 1994.

{}~\vfill
\end{document}